\documentclass[footinbib,aps,pra,reprint,superscriptaddress]{revtex4-1}

\usepackage{amsmath,amssymb,braket,upgreek,bm,verbatim}
\usepackage[absolute]{textpos}
\usepackage{physics, graphicx}
\usepackage{amssymb,amsmath}
\usepackage{graphicx,tabularx} 
\usepackage{xcolor}

\usepackage[absolute]{textpos}
\usepackage{multirow}
\definecolor{darkblue}{rgb}{0,0,0.5}

\DeclareMathOperator{\sF}{\mathcal{F}}
\usepackage{hyperref}
\hypersetup{
colorlinks=true,
linkcolor=black,
filecolor=blue,
citecolor=darkblue,  
urlcolor=black,
}

\begin{document}
\title{Generation and characterization of ultrabroadband polarization--frequency hyperentangled photons}

\author{Hsuan-Hao Lu}
\email{luh2@ornl.gov}
\affiliation{Quantum Information Science Section, Computational Sciences and Engineering Division, Oak Ridge National Laboratory, Oak Ridge, Tennessee 37831, USA}

\author{Muneer~Alshowkan}
\affiliation{Quantum Information Science Section, Computational Sciences and Engineering Division, Oak Ridge National Laboratory, Oak Ridge, Tennessee 37831, USA}

\author{Karthik V. Myilswamy}
\affiliation{School of Electrical and Computer Engineering and Purdue Quantum Science and Engineering Institute, Purdue University, West Lafayette, Indiana 47907, USA}

\author{Andrew M. Weiner}
\affiliation{School of Electrical and Computer Engineering and Purdue Quantum Science and Engineering Institute, Purdue University, West Lafayette, Indiana 47907, USA}

\author{Joseph M. Lukens}
\affiliation{Quantum Information Science Group, Oak Ridge National Laboratory, Oak Ridge, Tennessee 37831, USA}
\affiliation{Research Technology Office and Quantum Collaborative, Arizona State University, Tempe, Arizona 85287, USA}

\author{Nicholas~A.~Peters}
\affiliation{Quantum Information Science Section, Computational Sciences and Engineering Division, Oak Ridge National Laboratory, Oak Ridge, Tennessee 37831, USA}

\date{\today}

\maketitle

\begin{textblock}{13.3}(1.4,15)
\noindent\fontsize{7}{7}\selectfont \textcolor{black!30}{This manuscript has been co-authored by UT-Battelle, LLC, under contract DE-AC05-00OR22725 with the US Department of Energy (DOE). The US government retains and the publisher, by accepting the article for publication, acknowledges that the US government retains a nonexclusive, paid-up, irrevocable, worldwide license to publish or reproduce the published form of this manuscript, or allow others to do so, for US government purposes. DOE will provide public access to these results of federally sponsored research in accordance with the DOE Public Access Plan (http://energy.gov/downloads/doe-public-access-plan).}
\end{textblock}

\textbf{We generate ultrabroadband photon pairs entangled in both polarization and frequency bins through an 
all-waveguided Sagnac source
covering the entire optical C- and L-bands (1530--1625~nm). We perform comprehensive characterization of high-fidelity states in multiple dense wavelength-division multiplexed channels, achieving full tomography of effective four-qubit systems. 
Additionally, leveraging the inherent high dimensionality of frequency encoding and our electro-optic measurement approach, we demonstrate the scalability of our system to higher dimensions, reconstructing states in a 36-dimensional Hilbert space consisting of two polarization qubits and two frequency-bin qutrits. Our findings hold potential significance for quantum networking, particularly dense coding and entanglement distillation in wavelength-multiplexed quantum networks.
}
\smallskip

Photonic hyperentanglement typically describes two-photon states that exhibit simultaneous entanglement in multiple degrees of freedom (DoFs), e.g., orbital angular momentum, spatial mode, time-frequency, and polarization~\cite{Kwiat1997, Barreiro2005, Barreiro2008, Williams2017, Reimer2019, Imany2019, Ecker2021, Chen2022, Francesconi2023}. The expansion of the Hilbert space enables deterministic controlled operations between two DoFs within a single photon, showcasing significant potential for quantum communication protocols including dense coding~\cite{Barreiro2008, Williams2017} and single-copy entanglement distillation~\cite{Simon2002, Ecker2021}. Among the various exploitable DoFs, the polarization DoF has historically received extensive investigation, primarily due to readily available tools for state manipulation. On the other hand, time-frequency encoding stands out as a promising candidate due to its compatibility with established fiber-optic networks. Specifically, discrete frequency bins~\cite{Kues2019, Lu2019c}, a special case under the wider time-frequency umbrella, offer practical advantages such as straightforward multiplexing, parallel processing of multiple qubits, and the absence of nested interferometers that typically require active stabilization.

In this work, we present an all-waveguided, ultrabroadband hyperentangled source spanning polarization and frequency-bin DoFs. We perform the first full quantum state tomography (QST) of polarization and frequency-bin hyperentangled states, covering multiple dense wavelength division multiplexing channels across the optical C-band (1530--1565~nm) and L-band (1565--1625~nm) and extend measurements to high-dimensional Hilbert spaces through frequency-bin \emph{qudit} encodings. Through the serial application of polarization projections and electro-optic-based frequency mixing, our scheme can probe arbitrary bases in the complete two-photon polarization and frequency Hilbert space, thus facilitating full state reconstruction with no constraints on the ground truth state. Our procedure is experimentally demonstrated for up to 36-dimensional hyperentangled systems
, yet is in principle scalable to much higher dimensions. 
Overall, our source design and characterization techniques open new avenues for hyperentanglement generation and manipulation in these two important DoFs.


\begin{figure*}
\centering
\includegraphics[width=7in]{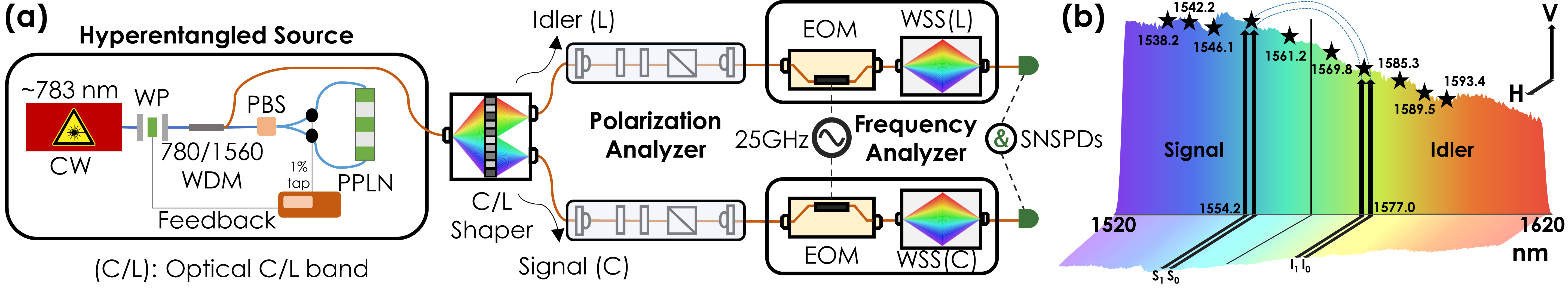}
\caption{(a) Experimental setup. (b) Conceptual diagram of the hyperentangled source. The background traces the SPDC spectrum measured on an optical spectrum analyzer (shown in linear scale)~\cite{Alshowkan2022}. The stars represent the five channel pairs characterized in this study. 
[CW: Continuous-wave laser. WP: liquid crystal waveplate. WDM: wavelength division multiplexer.
PBS: polarizing beamsplitter.
PPLN: periodically poled lithium niobate waveguide.
EOM: electro-optic phase modulator.
WSS: wavelength-selective switch.
SNSPD: superconducting nanowire detector.]}
\label{fig1}
\end{figure*}

Figure~\ref{fig1}(a) depicts the experimental setup, encompassing the hyperentangled photon source and two stages of state analyzers for each photon: one for polarization and one for frequency bins. We operate a continuous-wave laser around 783~nm and pump a 12~mm-long type-0 periodically poled lithium niobate (PPLN) ridge waveguide (AdvR) in a fiber Sagnac loop~\cite{Li2005, Fan2007, Vergyris2017, Alshowkan2022}. We employ a combination of a liquid crystal waveplate (WP; Thorlabs) and a fiber polarizing beamsplitter (PBS) to split the laser and coherently pump the waveguide from both directions. Using a 90-degree rotated fiber in one of the PBS outputs aligns pump photons and generated photon pairs in both directions to vertical polarization while traversing the PPLN waveguide. Upon recombination at the PBS and the 780/1560~nm wavelength-division multiplexer, the generated photon pairs are in the form of a polarization-entangled state $\ket{\Psi_P}\propto\alpha\ket{HH}+\beta\ket{VV}$, supporting a two-photon bandwidth of $\sim$18 THz, with the signal and idler photon covering more than the optical C- and L-band, respectively. The present source, which builds upon the design presented in Ref.~\cite{Alshowkan2022}, has previously shown high-fidelity polarization entanglement across 150 pairs of 25~GHz-wide channels. In this work, we introduce a notable stability improvement by tapping 1\% of the pump power in the Sagnac loop for active control of the WP. This allows us to maximize the polarization entanglement (i.e., attain $|\alpha|\approx|\beta|$) and counter real-time polarization drifts in the pump laser. 
With this addition, 31 independent polarization QST trials on the same 25~GHz-wide channel pair were found to yield a mean fidelity of 98.6\% with a standard deviation of only 0.3\% over the course of 18 hours. 


The broadband spectral coherence and energy correlation between signal and idler photons also provide a natural resource for investigating frequency-bin entanglement and higher-dimensional Hilbert spaces~\cite{Lu2018b, Lu2019c, Kues2019, Lu2022b, Chen2022}. Several techniques have been explored to create discrete frequency bins: one approach integrates resonant structures into the pair-generation process~\cite{Reimer2019, Imany2019, Lu2022b}; alternatively, continuous biphoton spectra can be shaped into bins using external cavities~\cite{Lu2018b, Lu2020b} or programmable frequency filters~\cite{Sandoval2019, Chen2022}. Here we adopt the last of these configurations by introducing a Fourier transform pulse shaper (Finisar Waveshaper 4000B) to carve out pairs of energy-correlated frequency bins: Fig.~\ref{fig1}(b) highlights a qubit example with two bins in the L-band ($I_0$ and $I_1$) and two in the C-band ($S_0$ and $S_1$). Each bin is designed 
with a width of 18~GHz and spaced 25~GHz apart. Additionally, the same pulse shaper splits the signal and idler photons into separate optical fibers for subsequent state characterization. Assuming entanglement in both DoFs, the ideal entangled two-photon state can be expressed as follows:
\vspace{-0.1in}
\begin{multline}
\label{Psi}
    \ket{\Psi_{PF}} = \ket{\Psi_P} \otimes \ket{\Psi_F}\\
    = (\alpha\ket{HH}+\beta\ket{VV})    \otimes \sum_{k=0}^{d-1} \gamma_k \ket{\omega_k^{(I)} \omega_{d-1-k}^{(S)}},
    \vspace{-0.1in}
\end{multline}
where $\omega_k^{(I)} = \omega_0^{(I)} + k\Delta\omega$ and $\omega_{d-1-k}^{(I)} = \omega_{d-1}^{(S)} - k\Delta\omega$ denote frequencies carrying photons chosen from the spectrum such that $\omega_0^{(I)} + \omega_{d-1}^{(S)} =\omega_p$ (the pump frequency).
The broadband nature resulting from the type-0 phase matching condition ideally yields $|\gamma_k|\approx\frac{1}{\sqrt{d}}$ for all bins of interest. Any nonuniformity can be rectified via the pulse shaper~\cite{Bernhard2013} (not required in our experiments). 

For a hyperentangled photon pair with $N$ DoFs, each having encoding levels $d_1, d_2, ..., d_N$, a complete state reconstruction typically requires $\mathcal{O}\left(\prod_{i=1}^{N} d_i^{4}\right)$ linearly independent local projections across these $N$ DoFs, as the minimum number of independent parameters describing a mixed state scales quadratically with the Hilbert space dimension~\cite{Bertlmann2008}. For instance, a total of $2^4\times3^4=1296$ measurements were performed in Ref.~\cite{Barreiro2005} to achieve full tomography of a $(d_1\otimes d_1)\otimes(d_2\otimes d_2) = (2\otimes2) \otimes (3\otimes3)$ system, which involved a pair of polarization qubits and orbital angular momentum qutrits.
An experimentally simpler approach involves QST of each DoF independently, using
semidefinite programming to compute a lower bound on the global state fidelity from information in reduced density matrices~\cite{Chen2020, Chen2022}. This approach reduces the number of measurements required for state characterization 
to $\mathcal{O}\left(\sum_{i=1}^{N} d_i^{4}\right)$ (a sum over DoFs rather than a product)
, particularly valuable when scaling to higher dimensions. However, it is crucial to note that these measurements alone cannot provide a complete characterization of hyperentangled states.

We follow the first approach by performing local projections simultaneously in both DoFs. For polarization tomography, we employ two motorized polarization analyzers comprising free-space collimators, a quarter-wave plate, a half-wave plate, and a polarizing beamsplitter, with a total throughput of $\sim$70\%. Frequency measurement is a more intricate process. Previous methods in the context of polarization--frequency hyperentanglement~\cite{Chen2022, Francesconi2023} have relied on Hong--Ou--Mandel interference to verify frequency entanglement and then infer the density matrix under specific state assumptions~\cite{Ramelow2009}. However, for slow (integrating) detectors, 
this approach is sensitive to two-dimensional frequency-bin entanglement only, where the signal and idler photons also share identical spectra~\cite{Lingaraju2019}. For more general state analysis applicable to high-dimensional and nondegenerate photons, we utilize an electro-optic phase modulator (EOM; EOspace) and wavelength-selective switch (WSS; Finisar) to implement the necessary projective measurements~\cite{Lu2018b, Sandoval2019, Lu2020b, Lu2022b} for the frequency DoF. 
It is crucial to execute the frequency-bin projection \emph{after} the polarization analyzer due to the polarization sensitivity of the phase modulation
; in this way, each EOM receives a fixed physical polarization state throughout and can provide consistent modulation and transmission efficiencies.

To achieve full reconstruction of the actual global state $\rho_{PF}$ with qubit frequency encoding [$d=2$ in Eq.~\eqref{Psi}], we employ a total of $16\times 8$ local projections across the two DoFs: i.e., 16 polarization projections~\cite{James2001} paired with 8 frequency-resolved measurements (4 projections in the Pauli $Z\otimes Z$ basis, and 4 in the $X\otimes X$ basis). In the $Z\otimes Z$ measurement, the EOMs are turned off and the two WSSs---one for the signal photon (C-band WSS) and one for the idler photon (L-band WSS)---demultiplex each photon by color, amounting to 
a $2\times2$ joint spectral intensity measurement. The $X\otimes X$ measurement is equivalent to preceding the $Z\otimes Z$ measurement with two parallel Hadamard operations, which we implement probabilistically by driving the EOMs at the bin spacing (25~GHz) with a modulation index of $1.435$~rad~\cite{Lu2020b}. While we could implement a more traditional tomographically complete measurement set in frequency bins, 
the strong correlations observed in these two mutually unbiased bases (MUBs) are sufficient to provide a clear entanglement witness~\cite{Coles2017} 
that, when combined with Bayesian inference, can be used to estimate the full quantum state with low uncertainty~\cite{Lu2018b}.
Significantly, this situation is possible with a completely uniform prior distribution---i.e., with no a priori restrictions on the form of the state itself---as Bayes' theorem automatically extracts all the information available in a given measurement set via its logical framework~\cite{MacKay2003}. We consequently leverage this feature to streamline the number of measurement settings for Bayesian QST~\cite{Blume2010,Lukens2020b} below.

\begin{figure}[!t]
\includegraphics[width=3.3in]{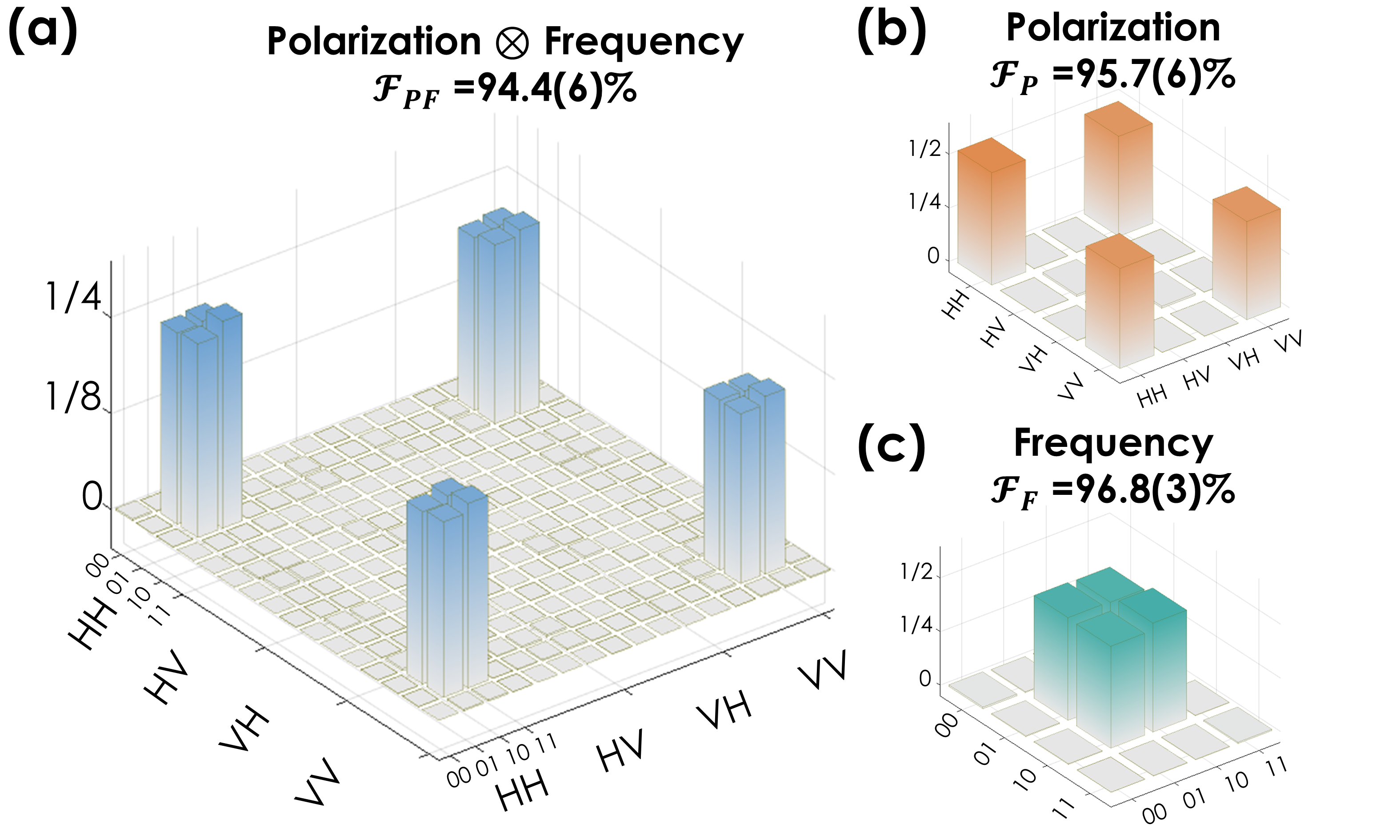}
\caption{Bayesian mean density matrices (real parts) of the $(2\otimes2)_P \otimes (2\otimes2)_F$ hyperentangled state [arrows in Fig.~\ref{fig1}(b)] and its reduced states in individual DoFs. The imaginary components (not shown) are less than 0.005 for all density matrices.}
\label{fig2}
\end{figure}

Figure~\ref{fig2}(a) presents the Bayesian mean density matrix $\rho_{PF}$ for a channel pair with signal and idler photons centered at 1554.2~nm and 1577.0~nm, respectively, obtained from 128 projections with 60~s integration time per point. Computing the fidelity of 1024 density matrix samples with the target state [Eq.~\eqref{Psi} with $\alpha=\beta$ and $\gamma_0=\gamma_1$], we find $\sF_{PF}=94.4(6)\%$---a value that simulations suggest is constrained by the number of coincidence events, rather than the quality of the states.
We also evaluate the reduced states in polarization ($\rho_P=\Tr_F \rho_{PF}$) and frequency ($\rho_F=\Tr_P \rho_{PF}$) by computing the respective partial traces. The density matrices and fidelities with respect to $\ket{\Psi_F}$ and $\ket{\Psi_P}$ in Eq.~\eqref{Psi} are presented in Fig.~\ref{fig2}(b,c). 
Quantitatively, we can lower- and upper-bound the distillable entanglement per DoF with the coherent information $I_C$ (maximized over one-way communication direction)~\cite{Devetak2005} and logarithmic negativity $E_\mathcal{N}$~\cite{Vidal2002}, respectively. From the QST data, we obtain the intervals $[I_C,E_\mathcal{N}] = [0.69(3), 0.936(9)]$~ebits for $\rho_{P}$ and  $[I_C,E_\mathcal{N}] = [0.76(2), 0.954(5)]$~ebits for $\rho_{F}$, confirming clear usable entanglement in both DoFs. 
Replicating the tomographic 
procedure for four additional channel pairs spanning the spectrum, denoted by stars in Fig.~\ref{fig1}(b), we measure the following hyperentangled state fidelities (counting outward from the spectral center), denoted as ordered pairs $(\mathcal{F}_{PF},\mathcal{F}_P,\mathcal{F}_F)$: $(93.3(7),94.5(6),95.9(4))\%$, $(93.3(7),94.5(6),96.1(4))\%$, $(93.7(8),94.8(7),96.7(3))\%$, and $(91.3(9),93.1(8),94.8(5))\%$.
Such a sampling indicates that true hyperentanglement persists across the bandwidth as expected.

\begin{figure}[b]
\includegraphics[width=3.3in]{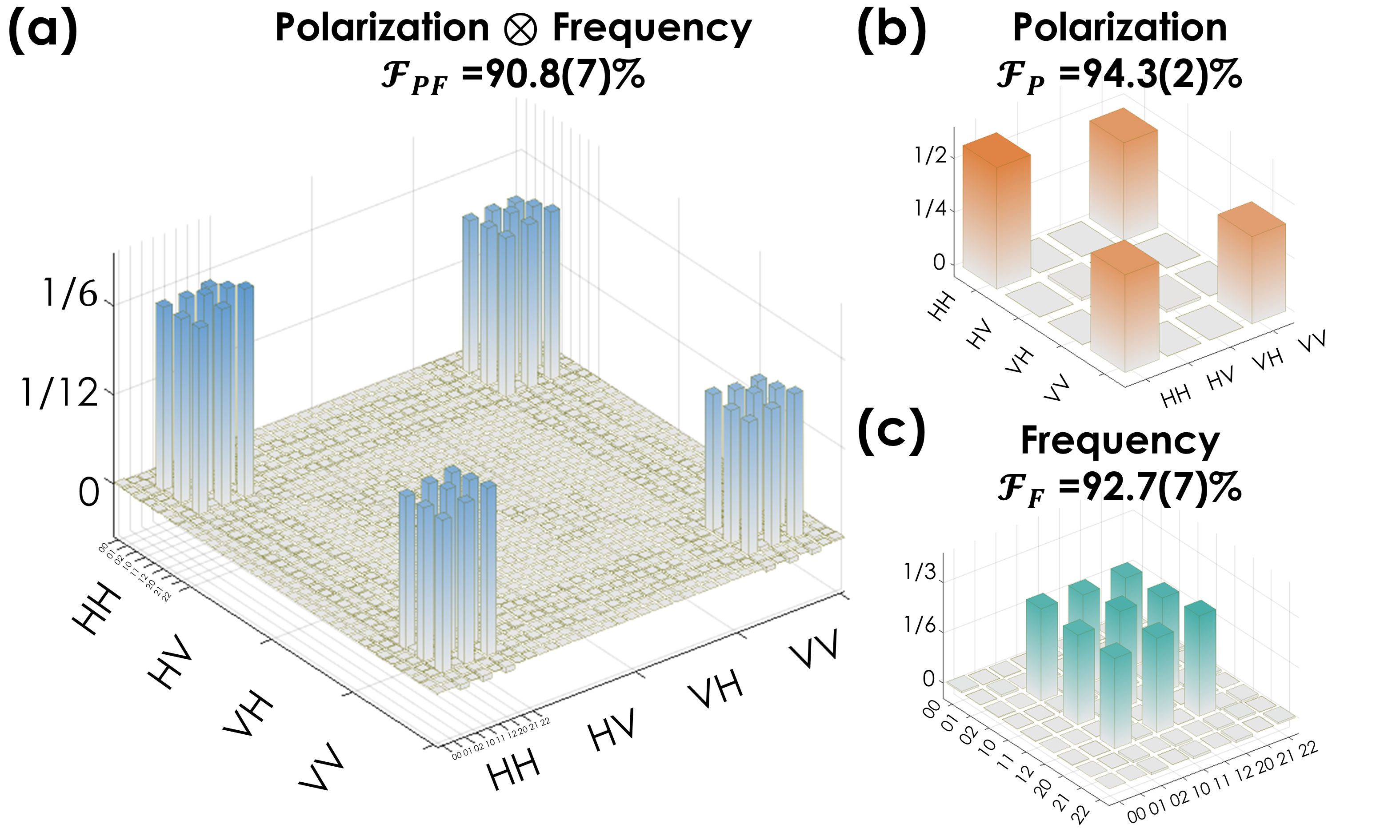}
\caption{Bayesian mean density matrices (real parts) of the $(2\otimes2)_P \otimes (3\otimes3)_F$ hyperentangled state and its reduced states in individual DoFs. The imaginary components (not shown) are less than 0.05 for all density matrices.}
\label{fig3}
\end{figure}

Given the broadband nature of the generated photon pairs, expanding the frequency dimensions 
is straightforward. 
For instance, to create frequency qutrits ($d=3$), we can simply consider three pairs of 25~GHz-spaced, 18~GHz-wide bins, ideally resulting in the state 
$\ket{\Psi_{PF}} = \ket{\Psi_P}\otimes \frac{1}{\sqrt{3}}\left(\ket{\omega_0^{(I)} \omega_2^{(S)}}+\ket{\omega_1^{(I)} \omega_1^{(S)}}+\ket{\omega_2^{(I)} \omega_0^{(S)}}\right)$. 
However, as the system's dimensionality increases, so does the number of measurements required for QST. In the interests of speed, we take advantage of the high degree of polarization entanglement and consider measurements in the $Z\otimes Z$ and $X\otimes X$ MUBs only (the same pair explored for frequency bins in the qubit example), thus reducing the number of polarization measurements from 16 to 8. 
To characterize high-dimensional frequency-bin entanglement, we leverage a novel method based on random measurements. This involves applying random phases with the pulse shaper and random frequency mixing with the EOMs, followed by computational-basis measurements~\cite{Lu2022b}. We consider a total of 720 measurements (60~s each): 8 polarization projections, 10 different random EOM and shaper settings (motivated by findings in~\cite{Lu2022b}), and $d\times d=9$ signal-idler frequency-bin combinations. Specifically, for each setting, the pulse shaper applies $2d = 6$ random spectral phases between $0$ and $2\pi$ to the aforementioned frequency bins, and both EOMs receive a sinusoidal voltage with amplitude $\delta$ chosen randomly between $0$ and $2.32$ radians~\cite{Lu2022b}. 
The resulting mean density matrix [Fig.~\ref{fig3}(a)] shows high-fidelity hyperentanglement: $\sF_{PF}=90.8(7)\%$ in the $(2\otimes2)_P \otimes (3\otimes3)_F$ system
. We again compute the reduced states in both DoFs [Fig.~\ref{fig3}(b,c)], with measured distillable entanglement intervals of $[I_C,E_\mathcal{N}] = [0.62(1),0.915(3)]$ and $[I_C,E_\mathcal{N}]=[1.04(4), 1.48(1)]$ ebits
for the polarization and frequency DoFs, respectively, to be compared with the maximum qubit limit of 1~ebit and qutrit limit of 
1.58~ebits.


Looking ahead, we see no immediate obstacles to generating even higher-dimensional hyperentangled states by expanding the frequency dimension, which in principle is capped only by
the ratio of total bandwidth to the bin spacing. 
Nevertheless, introducing more spectral content may degrade polarization entanglement due to increased sensitivity to polarization-mode dispersion (PMD)~\cite{Gordon2000}. This could introduce undesired polarization--frequency correlations that will ultimately depend on the specific fiber channel. To augment dimensionality without increasing bandwidth (and hence protecting the state from PMD impairments), narrower frequency spacings can be pursued instead, yet
it is important to note that commercial diffractive pulse shapers and WSSs usually have resolutions $\gtrsim$10 GHz. Ultimately, the optimal solution could involve a fully integrated version of the Sagnac source~\cite{Untitled}, supporting polarization diversity and allowing direct definition of frequency bins through the optical resonances of microrings. 
Importantly, in any case where PMD limits the usable frequency-bin dimension for an \emph{individual} state $\ket{\Psi_{PF}}$, the remaining bandwidth can still be leveraged for \emph{parallelization}, in which the output is sliced into subbands that are each sufficiently narrow to evade PMD degradation but collectively utilize the entire band. Such a source could be used for wavelength-multiplexed entanglement distribution, but where the intra-channel frequency-bin entanglement carries explicit quantum information as well as the polarization DoF.

Finally, the ability to \emph{manipulate} the expanded Hilbert space is the key to fully harnessing the potential of hyperentangled states. For example, 
controlled unitaries between polarization and frequency DoFs~\cite{Xu2023} will be valuable for implementing hyperentanglement-based versions of
protocols such as dense coding~\cite{Barreiro2008}, superdense teleportation~\cite{Graham2015, Chapman2020}, and entanglement distillation~\cite{Simon2002, Ecker2021}. 
However, the EOMs utilized in our measurement scheme 
are polarization-sensitive, which limits their suitability for certain applications in this context. While we can evade this restriction in QST by placing the EOMs after polarization projections, such a simplification will not be feasible for general quantum operations in the joint Hilbert space. Further advances in polarization-diverse/insensitive frequency modulation techniques~\cite{Ikuta2018, Sandoval2019} will therefore prove valuable to fully utilize such states in multidimensional quantum information processing. 

\section*{Acknowledgments}
The authors would like to thank B.~T. Kirby for valuable discussions. This work was performed in part at Oak Ridge National Laboratory, operated by UT-Battelle for the U.S. Department of energy under contract no. DE-AC05-00OR22725. Funding was provided by the U.S. Department of Energy, Office of Science, Advanced Scientific Computing Research (Field work Proposals ERKJ378, ERKJ353), and the National Science Foundation
(2034019-ECCS).

%

\end{document}